# Quantum random flip-flop based on random photon emitter and its applications in over-Turing computers, cryptography, signal processing and generation

Mario Stipčević[1]
[1]Ruđer Bošković Institute, Bijenička 54, 10002 Zagreb, Croatia
*corresponding author e-mail: Mario.Stipcevic@irb.hr
Date: August 28, 2013

***Abstract*— In this paper we propose, experimentally realize and study possible applications of a new type of logic element: random flip-flop. By definition it operates similarly to the conventional flip-flop except that it functions with probability of 1/2 otherwise it does nothing. We demonstrate one practical realization of the random flip-flop based on optical quantum random number generator and discuss possible usages of such a device in computers, cryptographic hardware and testing equipment.**

## 1. Introduction

Practically all contemporary computers are constructed from deterministic logic circuits (gates, flip-flops, etc.) and are therefore equivalent to a Turing machine [1]. However, it has been realized that a more powerful notion of a computer, so called probabilistic Turing machine [2], [3], can execute *randomized algorithms* [4] which offer either greater speed or greater simplicity (or both) than their deterministic counterparts and on top of they may be more resilient to errors which may occur during the calculation process. For example, one of the earliest known randomized algorithms, the Rabin-Miller primality test [5-6] efficiently (much faster than any known deterministic algorithm [7]) estimates with a high probability whether a given large number is prime or not which is highly useful in cryptography. Examples of use of probabilistic Turing machine include numerical simulations, Monte Carlo calculations, PSI factor research and gambling automata. And last but not least the cryptography, as one of the cornerstones of online trade and internet security, depends on local availability of freshly generated, unguessable random numbers at each communication endpoint.

Probabilistic Turing machine can be thought of as an ordinary Turing machine complemented with a random number generator although in its operation it deals both with random decisions and random numbers. Having in mind importance of random numbers or random decisions for countless applications it is quite astonishing that modern computer processors do not contain a standardized randomness source and a set of machine language instructions dealing with random decisions and/or random numbers. Instead, established state of the art is to use pseudo-random formulae seeded by user supplied data or with a handful of bits collected in various ways from the computer hardware (hard disc access times, internal clocks/interrupts, temperature sensors, keystrokes, etc.) in order to generate long pseudo-random numbers which are then used as a substitute for true random numbers.

Modern Turing machines are built from deterministic sequential logic (e.g. the NAND gate) and non-sequential logic circuits (e.g. a flip-flop). Here we propose a single, new elementary non-sequential logic building block: *the random flip-flop* which can be used, among many other applications, to build a probabilistic Turing machine. Random flip-flop may be useful in many other applications such as signal processing, cryptography, laboratory equipment, analog computing, etc. In this study we will define, describe a few possible experimental realizations of the random flip-flop and discuss examples of its usage which extend over both digital and analog applications.

**2. Definition of the random flip-flop**

Conventional edge-triggered D-type flip-flop (DFF) shown in Fig. 1a) is an important building element of logic circuits, microprocessors, computers etc., usually used as a one bit data storage or as a frequency divider-by-two. The "edge-triggered" action means that flip-flop performs its action when clock pulse input (CP) changes state from LOW (logic 0) to HIGH (logic 1), that is upon the "rising edge" of the input pulse. Flip-flop does not care about pulse duration or the falling edge at the clock input (CP). In this work we will only refer to edge-triggered flip-flops and therefore this characterization will be dropped henceforth. Although there are several types of flip-flops in use (D, T, RS, JK, etc.) the D-type is universal in the sense that it can most easily mimic all other types of flip-flops. In principle, it can be realized from basic logic gates [8]. D-type flip-flop operates in the following way. With each occurrence of the rising edge at the input CP (clock pulse), the logic level present at the data input D is transferred to the output Q. The transfer takes some propagation time, usually denoted by $t_{CPQ}$. After the transfer is finished the value of Q is frozen i.e. becomes independent of the subsequent changes at the input D until occurrence the next rising edge at the clock input. This operation is illustrated by an example timing diagram shown in Fig. 1b). By definition, the D-type flip-flop is a *deterministic* device. It means that under the same input conditions it will always give the same output. Of course, conventional logic elements such as flip-flops, AND gates, OR gates etc. are designed specifically to behave in a deterministic fashion in order to ensure stable and predictable operation of logic circuits, computers, etc. Although in practical devices problem known as "metastability" [9] occurs with a small probability, it can be minimized by careful design of the flip-flop, by stacking flip-flops or by avoiding critical operating conditions. This elusive effect is of no interest here.

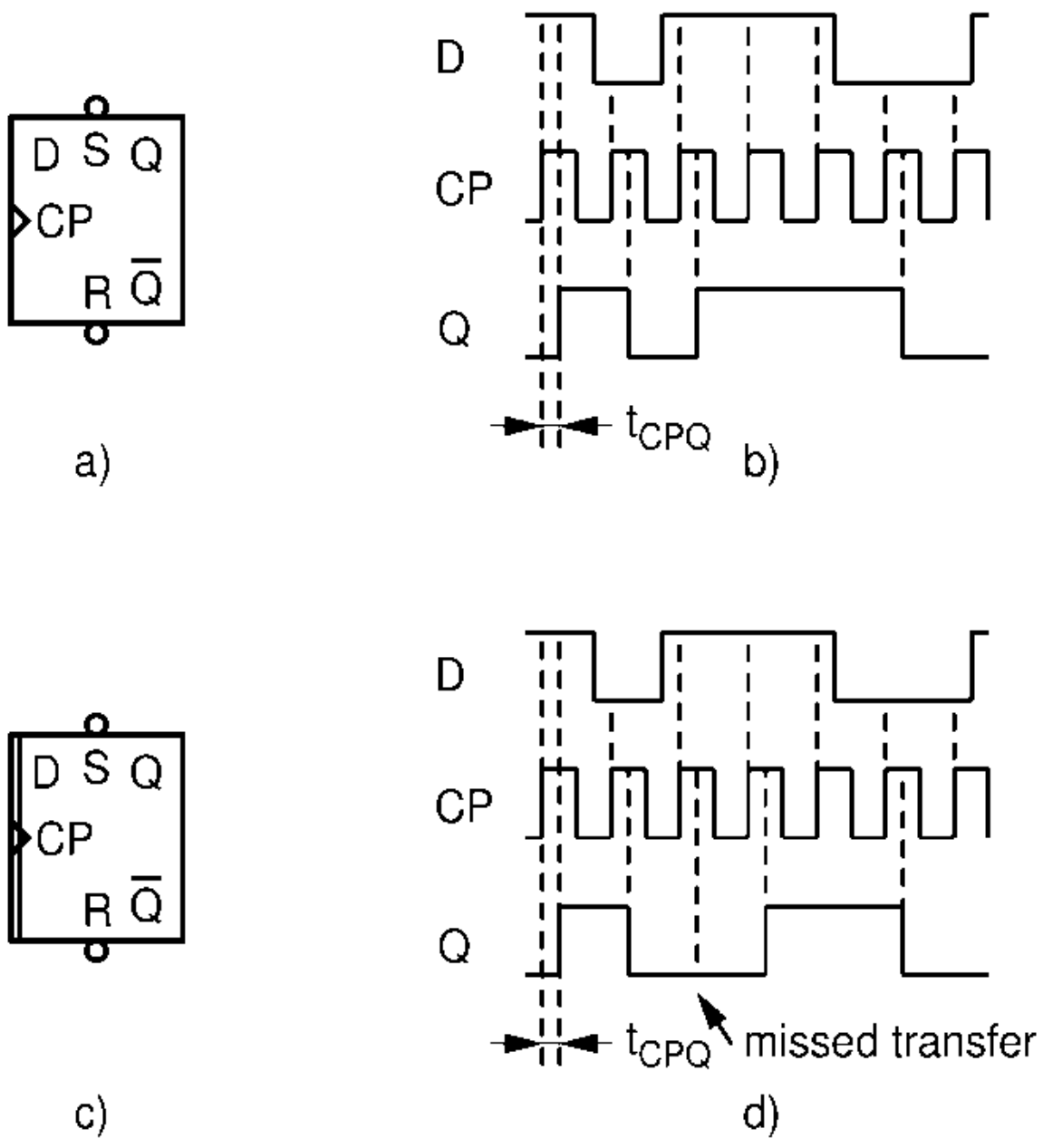

**Figure 1.** The conventional D-type flip-flop (DFF) symbol (a) and an example of its input-output waveforms (b). The D-type random flip-flop (DRFF) symbol (c) and an example input-output waveforms (d). The $t_p$ is propagation delay between the clock pulse input CP and the outputs Q and $\overline{Q}$.

As a contrast to deterministic flip-flops we here introduce a family of *random flip-flops* (RFF). Each deterministic type of flip-flop has its random counterpart. For example, by definition, the D-type random flip-flop (DRFF) (symbol shown in Fig. 1c)), operates the same as the conventional D-type flip-flop with only difference that upon the clock pulse, state of the data input D is transferred to the output Q with probability of ½, randomly. An example of the input-output diagram of DRFF is shown in Fig. 1d). We have adopted the usual convention that the state of the output changes with positive going edge of the clock pulse. Another type of random flip-flop of particular convenience for this presentation is T-type (TRFF). Deterministic T-type flip-flop (TFF) toggles output state with each clock if input T is held at logical 1 while if T=0 the output is left unchanged. By definition, TRFF is different from TFF only in that if T=1 then each clock pulse reverses the output state with probability ½. In both TFF and TRFF clock input is disabled while T=0. Note that the two definitions (of DRFF and TRFF) are mutually compliant in that whatever a given type of deterministic flip-flop does upon receiving a clock pulse, its random counterpart does randomly with probability of ½, and that in the same manner the definition could be extended to all other types of flip-flops. A RFF optionally features inverted output $\overline{Q}$ which is a binary complement of the output Q. Besides random action of the clock input CP, all random flip-flops optionally feature deterministic Set (S) and Reset (R) inputs or their inverted versions ($\overline{S},\overline{R}$), which unconditionally *set* (to HIGH) or *reset* (to LOW) the output Q. Although the two random flip-flops can be emulated by each other as shown in Fig. 2, it is sometimes more convenient to realize given circuit by use of a particular type. Interestingly, the same emulation schemae are true for deterministic counterparts (DFF and TFF).

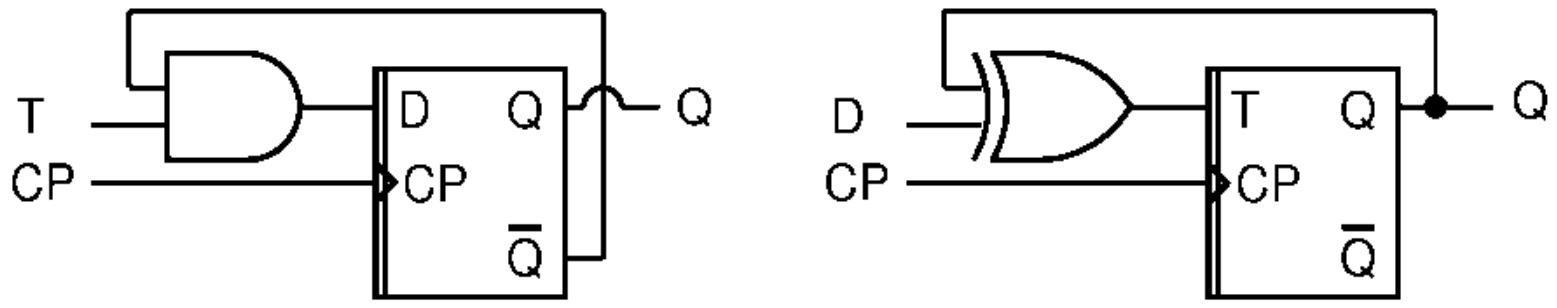

**Figure 2.** Emulations of DRFF and TRFF by each other. These exact same schematic diagrams hold also for deterministic counterparts.

Of course, one could consider a random flip-flop with the action probability different from 1/2 but the probability of 1/2 is the most convenient for general applications and it will be shown how other probabilities can be achieved (with arbitrary precision) by use of multiple RFF's.

## 3. Pseudo random flip-flop

By definition random flip-flop is a non-deterministic device, however in some cases pseudorandom behavior can be tolerated or perhaps preferred. If nothing else, pseudo-random flip-flop (PRFF) is interesting because it can be built from standard logic circuits and requires a little of resources. An example of a D-type pseudo-random flip-flop (DPRFF) which uses linear feedback shift register (LFSR) [10-11] is shown in Fig. 3.

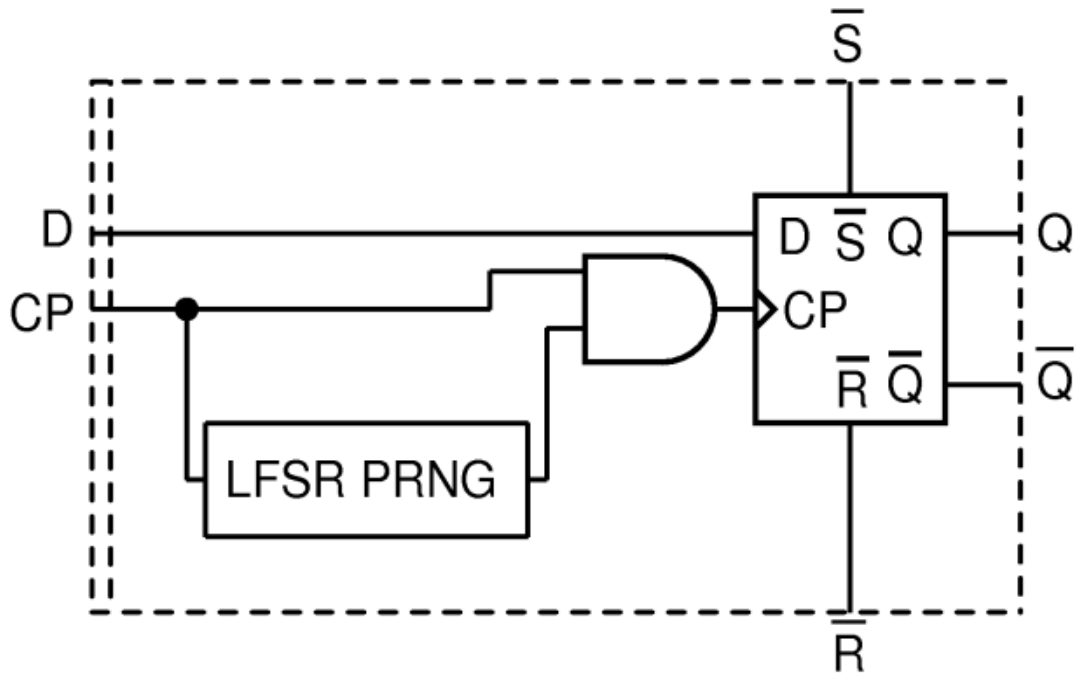

**Figure 3.** Pseudo-random flip-flop (PRFF) with LFSR.

It is generally straightforward to make PRFF with a constant delay time, a feature which is highly desirable but as we shall see not easy to obtain with sources of true randomness. Additional bonus of using LFSR is that the total propagation time of the RFF is constant and independent of the length of the LFSR so highly complex LFSRs can be used at no cost in speed. The propagation time of the resulting PRFF is very small: just about twice the propagation time of the DFF employed. However one has to bare in mind that pseudo-random flip-flop is in fact deterministic device, built only from deterministic logic elements and therefore its output is predictable. This compromises its usability in cryptographic and some other applications, as will be discussed.

## 4. Practical realization of an optical quantum random flip-flop

Quite generally, a random flip-flop must contain some source of randomness and a few deterministic logic circuits. There are two important considerations in conceiving a practical RFF.

First, the action of a RFF must be truly random and independent of all other RFFs used in the circuit.

Second, the propagation delay between clock pulse and the output(s), $t_{CPQ}$, must be constant, that is it may not vary from clock to clock *and* it should be the same for all RFFs used in a given circuit. The second requirement is important because otherwise RFF would not be suited for operation in a clocked environment and it would be hard to synchronize operation of several RFFs.

It is not easy to fulfill both requirements at the same time. At the current state of the art the best way to produce true randomness is to measure some fundamentally random physical phenomenon. But such phenomena generally appear at random times or at well defined times but with probability smaller than unity, so there is no guarantee that an event will be available at the time of request (clock). The solution to that can be found in buffering which would allow that a previously measured state is stored in a piece of memory (buffer) and waits to be used at the time of request. If this is the case we say that RFF is operating in *retarded* mode because its operation is determined by event(s) in the past. Esoterically, some special applications such as loop-hole free Bell inequality test, require that the course of action of the RFF is decided completely in the future of the clock [12]. This *advanced* mode is even harder to achieve (probably impossible with 100% efficiency), but for most applications the operation mode is of no importance.

Probably the simplest way to realize a RFF if by use of the constant propagation-delay physical random number generator (RNG) elaborated in [13] whose output would be used to control random behavior of the RFF. The randomness source in this RNG is a *random pulse train* (RPT) - a sequence of short logic pulses of constant height (amplitude) and width (duration) wherein waiting times $t_i$ between positive-going edges of subsequent pulses follow exponential probability density function, illustrated in Fig. 4. RPT is a very important concept in this presentation. It is characterized by its mean frequency $f$ and it is implicitly assumed that the pulse width is much smaller that $1/f$.

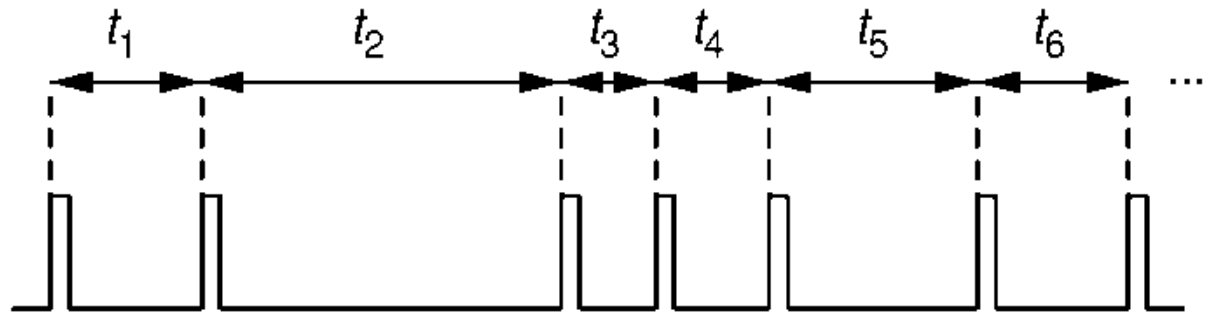

**Figure 4.** A random pulse train (RPT) consists of logic pulses of equal width and height. By definition, separation times $t_i$ between (positive-going edges of) adjacent pulses follow exponential probability density function.

The D-type RFF can now be constructed as shown in Fig. 5. The internal RNG consists of a constant-mean-frequency random pulse train generator RPG followed by a special extraction circuit consisting of two flip-flops FF1 and FF2. The T-type flip-flop FF1 toggles its state under influence of RPG at random times such that its output Q spends equal amount of time in either state (HIGH, LOW). When a clock CP arrives, the flip-flop FF2 produces a short pulse if and only if its output D is HIGH, that is with probability of exactly ½. It is shown in [13] that when the time between subsequent clock pulses is larger than the mean period of random events then the sampled state asymptotically becomes truly random as ratio of the two periods goes to infinity. The fastest convergence to true randomness is achieved for exponential random pulse generator.

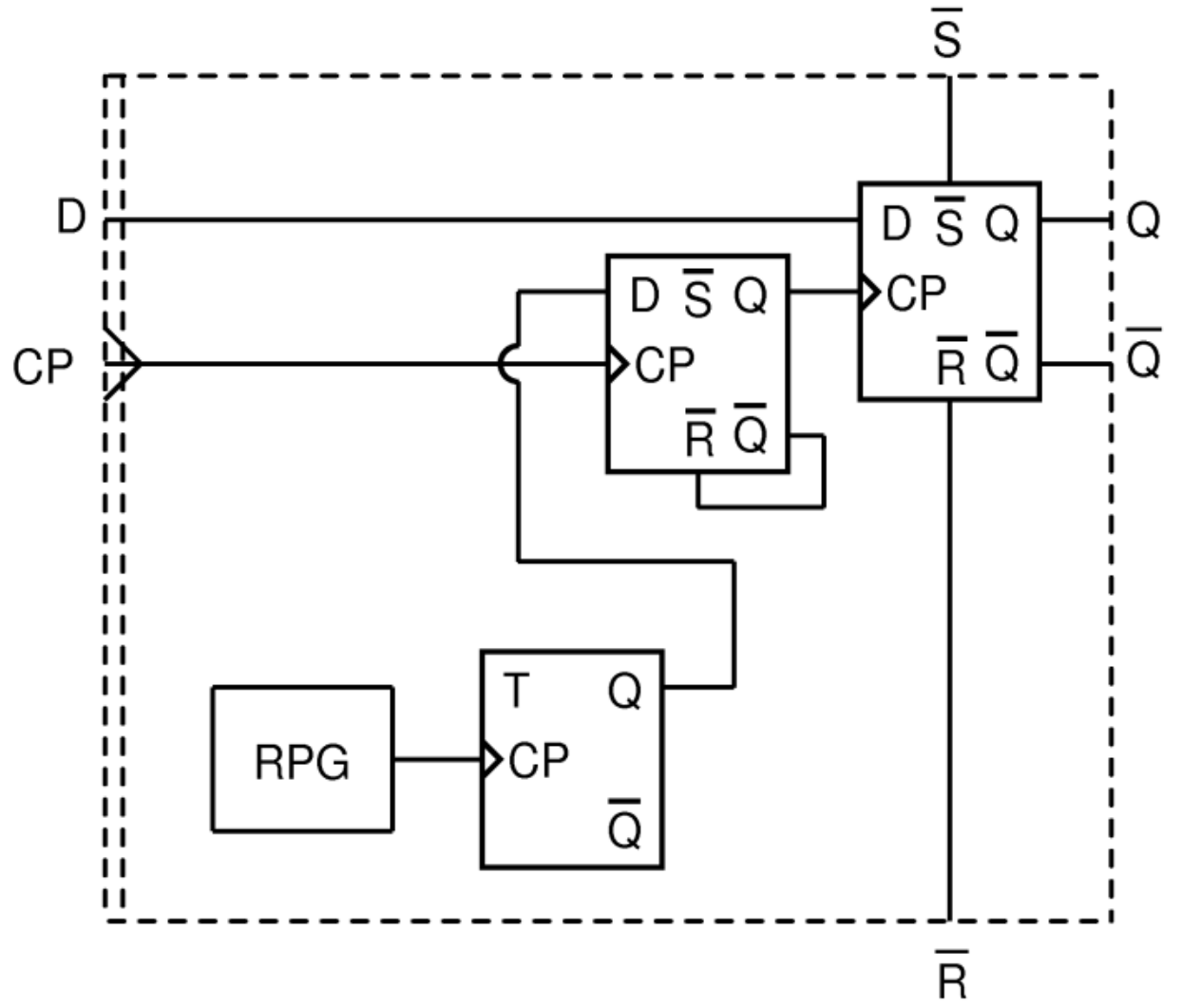

**Figure 5.** One possible realization of the random flip-flop by using a random pulse generator (RPG).

An optical RPG whose randomness principle relies on laws of quantum physics is described in [14]. It consists of a stationary source of random light which shines onto single photon detector which, upon each successful detection of a photon, generates one logic pulse of a constant width and height. For the light source one would ideally use a single photon light emitting diode (SPLED) which consists of one or more independent, electrically pumped quantum dots each of which can emit only one photon at a time. Ensemble of many such (at least partially independent and at least partially random) sources is a very good approximation of a Poissonian random source of single photons. Toshiba has first developed SPLED operating at cryogenic temperatures in 2002 [15-16] and recently demonstrated SPLED that operates at room temperature [17], however these remain unavailable to other researchers up to date. Similar technology of a single photon source based on quantum dots in a Fabry-Perot cavities has been described in [18] but with the same caveat.

Below we describe one possible practical realization of a RFF that could be built by widely available technology. For practical constructions one can resort to ordinary GaAs light emitting diode (LED) operated in a stationary, constant-power mode. One can roughly estimate the coherence time of such a source as $\tau_{cohr} = \lambda^2 /(c\Delta\lambda)$ which is on the order of few tens of femtoseconds for most LEDs that emit in the visible range. As long as the coherence time is much smaller than the smallest clock period the photon-pair time interval statistics obeys a nearly perfect exponential probability density function. In fact much more correlations among bits will come from imperfections in the photon detector such as dead time and afterpulsing because dead time and afterpulse lifetime are typically many orders of magnitude longer than the coherence time. But fortunately, as explained in [13], any short range correlations present in the RPG will have a vanishingly small effect on randomness of bits in the limit of low enough clock rate, specifically when the clock period is sufficiently long with respect to both dead time and afterpulse lifetime. One possible realization of DRFF using an optical RPT generator is shown in Fig. 6. The LED source, optionally attenuated by a dispersive (non-fluorescent) filter DF, shines upon the single photon detector SPD which produce random pulses. The DF filter is only supposed to remove a portion of photons without creating correlations in photon arrival times to the detector. In the rest of the paper we will present a collection of most interesting applications of random flip-flops.

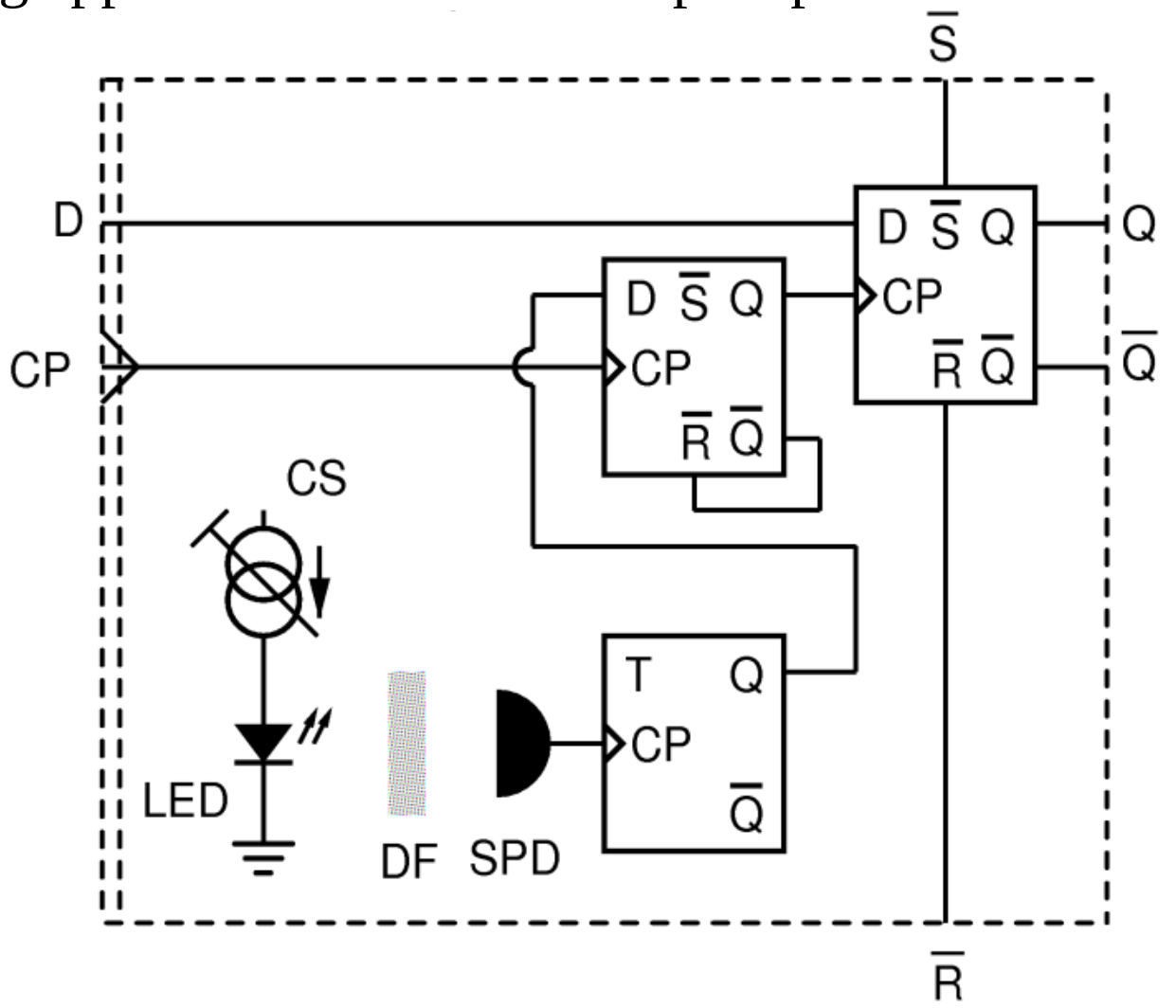

**Figure 6.** A practical constant delay D-type random flip-flop (DRFF) utilizing an optical random pulse train generator made with: 1) a stationary random photon source (LED supplied by adjustable current source CS); 2) dispersive optical filter (DF); and 3) a single photon detector (SPD).

## 5. Random number generator

Probably the simplest and most natural application of the random flip-flop is the random bit generator (RBG), a physical device that produces one random bit upon a request. By definition of a RBG, every generated bit must not have any correlation to the pool of previously generated bits *and* probability of zeros $p_0$ must be equal to the probability of ones $p_1$. Figure 7 shows two equivalent random bit generators realized with DRFF (left) and TRFF (right). Each rising edge at the CP input will trigger a generation of a fresh new random bit at the Data output synchronously with the Strobe's rising edge. By definition of RFF the sequence of random bits is truly random and there is no correlation of any sort among bits. A random number generator that produces multiple random bits per single clock can be obtained by parallelizing desired number of bit generators.

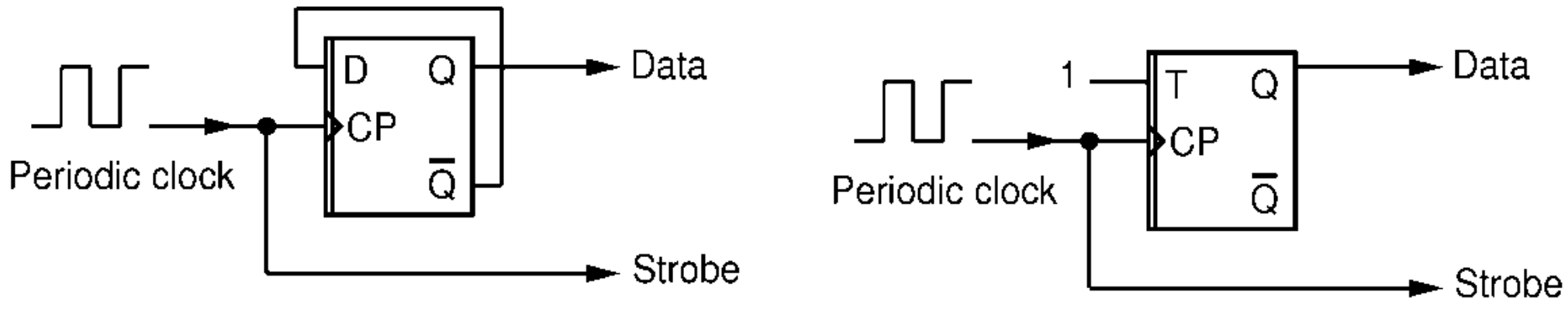

**Figure 7.** Stream random number generator. Shown are equivalent realizations with DRFF (left) and TRFF (right). Each rising edge of the periodic clock generates a fresh new random bit at the Data output synchronously with the Strobe's rising edge.

Having in mind possible miniaturization of RFFs to a chip level, random number generators of this sort can be particularly useful in cryptographic hardware (smart cards, password tokens, cell phones, computers,...) with a purpose of secure generation of keys, passwords, challenge-response data, quantum cryptography [19].

## 6. Randomness preserving frequency dividers

Dividing periodic frequency of periodic signals can be performed by various well known techniques [20-22]. For example, a single TFF will divide by two frequency of its clock signal while division by $n$ can be made by a counter-to-$n$. On the other hand, direct division of frequency of random pulse trains in such a way that randomness is preserved wasn't feasible so far. However, randomness preserving division can be made with random flip-flops. A random divider by 2, shown in Fig. 8a, omits randomly on average every second clock pulse. If an RPT of average frequency $f$ is brought to the input CP, an RPT of frequency $f/2_R$ will appear at the output Q. The subscript "R" denotes "random" division action. Such division preserves exponential distribution of pulse waiting times. Stacking two DRFFs as shown in Fig, 8b one achieves division by $4_R$ etc. A surprising possibility of dividing frequency by 4 with a single DRFF is shown in Fig 8c, but this division is composed of first dividing by 2 randomly and then by 2 deterministically, which we denote as $1/2_R/2_D$ where subscript "D" stands for "deterministic".

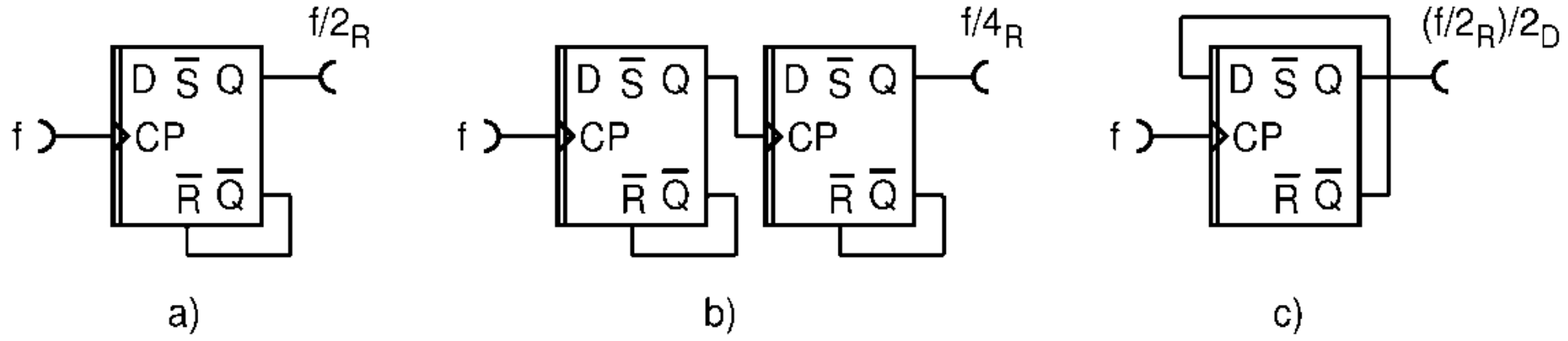

**Figure 8.** Basic dividers with RFFs: a) random divider by 2; b) random divider by 4; c) divider which performs random division by 2 followed by deterministic division by 2.

Although dividers in Figs. 8b) and 8c) both divide input frequency by 4 their action is not the same because the first divider preserves exponential distribution whereas the second converts exponential distribution into Erlang distribution with shape parameter $n$=2. Generally, a RPT with mean frequency $f$ deterministically divided by $n$ will result in another random pulse train with mean frequency of $f/n$ but its distribution will be Erlang distribution with shape parameter $n$:

$$\text{Erlang}(t, f, n) = \frac{f^n t^{n-1} e^{-ft}}{\Gamma(n)} \qquad (1)$$

which for large $n$ tends to the Normal distribution due to the central limit theorem. The Erlang distribution for $n$ = 2 which corresponds to waiting times of an exponential RPT divided by $2_D$ is shown in Fig. 9. A random division of an Erlang($t$, $f$=1, $n$=2) pulse train by factors $2_R$, $4_R$ and $16_R$ (normalized to unity frequency). The conclusion is that deterministic division by $n$ drives the output signal away from randomness and exponential distribution whereas random division by *n* drives *any* initial distribution asymptotically closer to the exponential p.d.f., that is to the highest possible randomness. In that respect random frequency division and synthesis of RPTs (Sect. 9) as well as RPT computing (Sect. 11) are stable, self-healing processes which tend to diminish effects of imperfections in hardware.

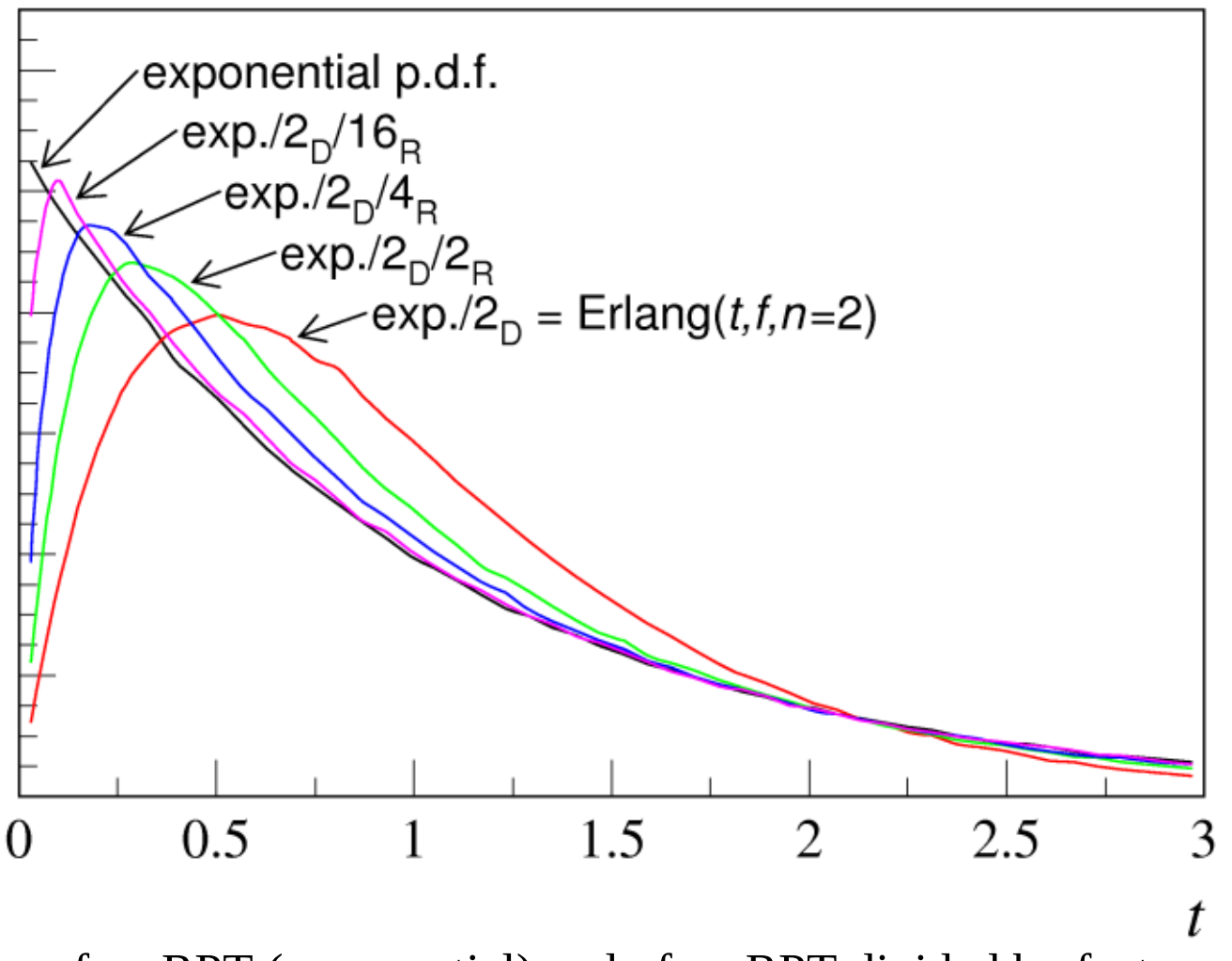

**Figure 9.** Waiting time distributions of an RPT (exponential) and of an RPT divided by factors: $2_D$, $2_D2_R$, $2_D4_R$ and $2_D16_R$.

Note that in the special case of dividing a RPT by $2_D$ (for example by means of a single deterministic flip-flop) no information is lost: every edge (positive-going or negative-going) represents one pulse from the original RPT and therefore it is possible to restore the original RPT form the divided one. This can be done by a deterministic multiplier by $2_D$, such as the one shown in Fig. 10, which upon every edge (rising or falling) generates one short pulse of width equal to the propagation delay of the non-inverting buffer. On the contrary, when a RPT is randomly divided by $2_R$ then half of the information (ever second pulse, on average) is lost.

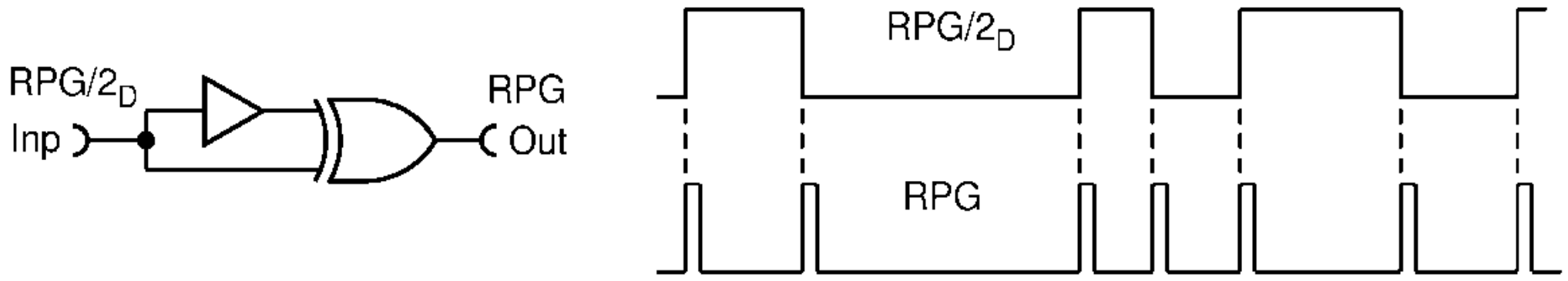

**Figure 10.** Deterministic multiplier by two: frequency of any input signal will appear multiplied by 2 at the output. The input and output waveforms are shown on the right. The output pulse width equals to propagation delay of the non-inverting buffer.

Later we will discuss other approaches to division and synthesis of frequency of random pulse trains.

## 7. Random counters

The $1/2_R/2_D$ frequency divider explained above (Fig. 8c) is equivalent of the random number generator in Fig. 7 left and can be considered as a single bit random counter. Indeed, usual approach to frequency division is by use of counters. Here we explore an equivalent of a deterministic synchronous 4-bit counter realized with TRFFs rather than with TFFs, shown in Fig. 11. This circuit has a rich set of complex properties.

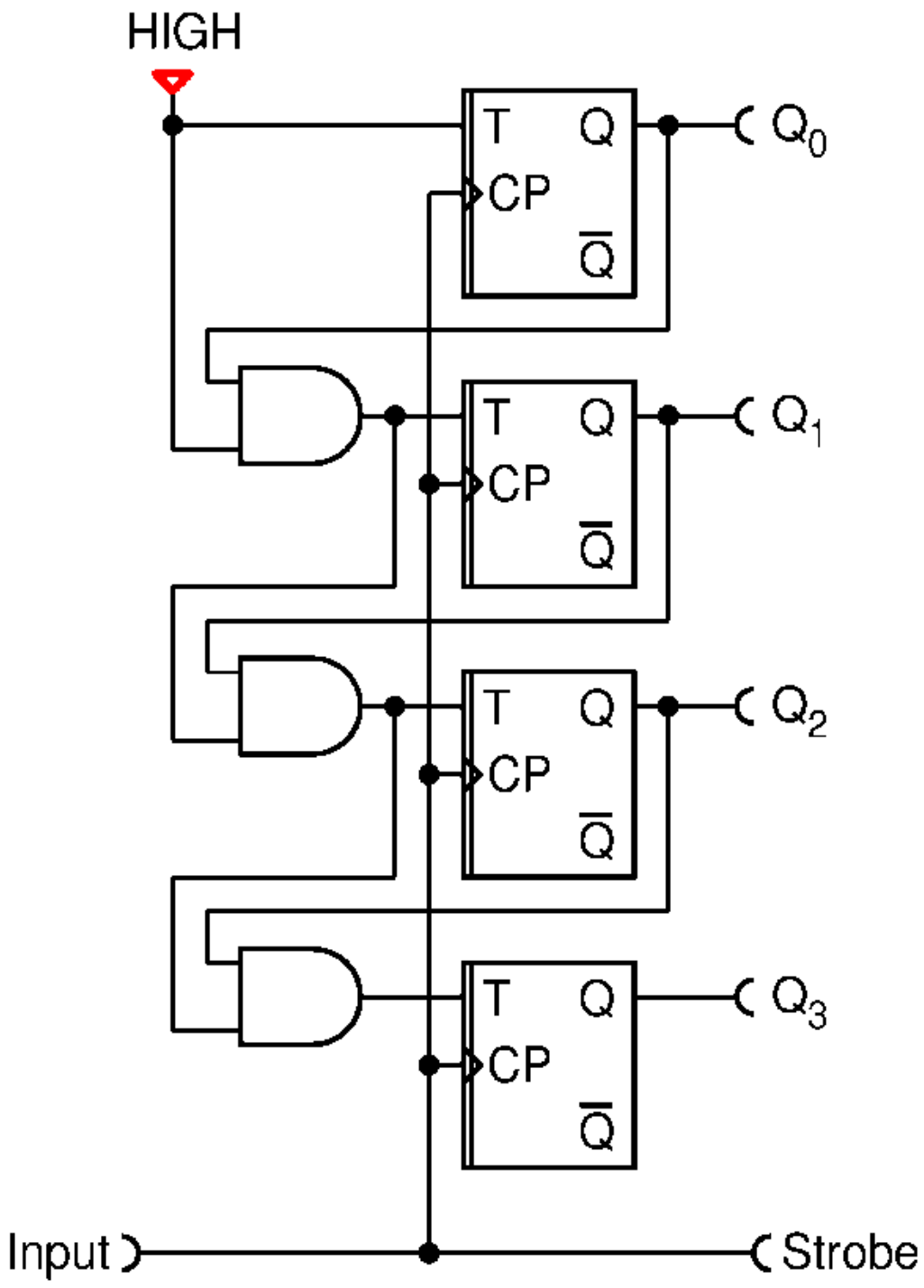

**Figure 11.** Synchronous 4-bit random counter. With each clock pulse the counter randomly advances up similar to random walker on a line.

The counter has 16 possible states which can be represented by a binary number $\{Q_3,Q_2,Q_1,Q_0\}$ whose value is in the range from 0 to 15. Unlike its deterministic which counts from 0 to 15, then falls back to 0 etc., the output value of the random counter behaves like a random walker that wonders around the mean value of 15/2. The binary numbers are random draws from the Binomial distribution with $n$ = 15 trials and trial success probability $p$ = ½. It can be used as a generator of such values, but at the moment it is not known how to make a random counter up to an arbitrary integer number $n$.

When considered individually, each of the outputs $Q_i$ (taken in synchronization with the Strobe) yields a random number sequence with zero bias but serial autocorrelation coefficient (defined for example in [23]) with a lag $k$ equal to: $a_k(Q_i) = (1\text{-}2^{-i})^k$. The coefficients obey geometric law with respect to the lag.

If Input is fed by a RPT, each output $Q_i$ gives another RPT. The mean frequency of $i$-th output ($Q_i$) is will be equal to the input mean frequency divided randomly by $2^{i+1}$ and then deterministically by 2 (in our notation: $f(Q_i) = f_{in}/(2^{i+1})_R/2_D$ ) but RPTs appearing at outputs $Q_0$-$Q_3$ are not statistically independent and cannot be combined to obtain RPTs of frequencies other than those. Another approach to frequency division, by random multiplication with a constant number less than 1, will be discussed in the next two sections.

## 8. Randomness preserving frequency multiplier

Another approach to lowering of the frequency of exponential random pulse trains (RPTs) while maintaining the exact exponential distribution of waiting time intervals is to multiply the frequency by a number smaller than one, by use of so called *random multiplier circuit*. This approach generally requires more resources than division by random counters but allows one to obtain finer gradations of the final frequency. The simplest random multiplier is the "$\mu$-multiplier", shown in the Fig. 12.

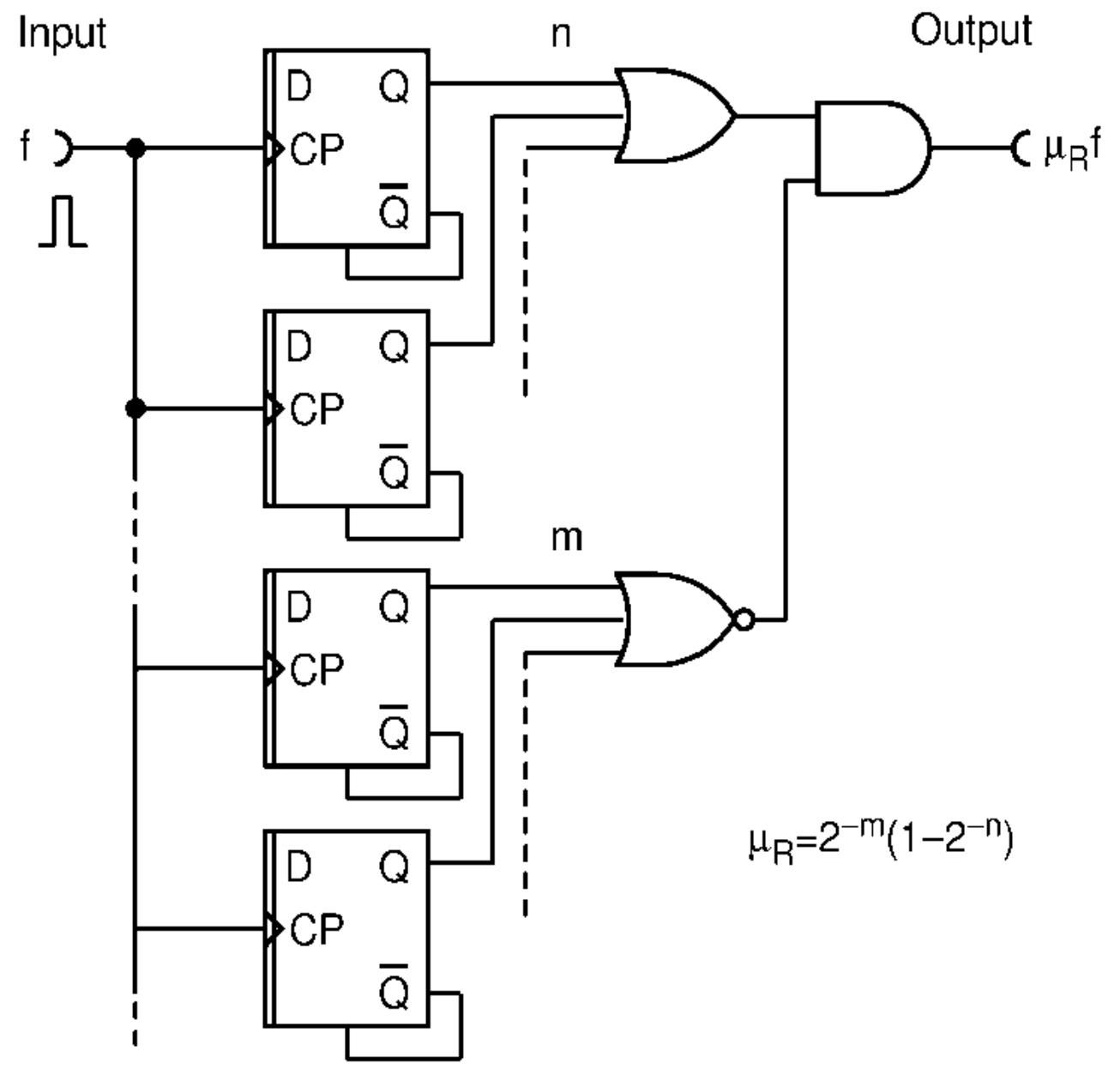

**Figure 12.** Random frequency multiplier. The frequency *f* of an input signal is random-multiplied by a factor $\mu_R = 2^{-m}(1\text{-}2^{-n})$, with conventions that if $m = 0$ then output of the NOR gate is held HIGH whereas if $n = 0$ then output of the OR gate is held HIGH. If input stream is an RPT then the output stream is also an RPT.

The frequency *f* of an input signal is random-multiplied by a factor $\mu_R = 2^{-m}(1\text{-}2^{-n})$, with conventions that if $m = 0$ then output of the NOR gate is held HIGH whereas if $n = 0$ then output of the OR gate is held HIGH.

| $N$ | $2^N$ | $k$ |
|---|---|---|
| 2 | 4 | 1 2 3 |
| 3 | 8 | 1 2 3 4 6 7 |
| 4 | 16 | 1 2 3 4 6 7 8 12 14 15 |
| 5 | 32 | 1 2 3 4 6 7 8 12 14 15 16 24 28 30 31 |
| 6 | 64 | 1 2 3 4 6 7 8 12 14 15 16 24 28 30 31 32 48 56 60 62 63 |
| 7 | 128 | 1 2 3 4 6 7 8 12 14 15 16 24 28 30 31 32 48 56 60 62 63 64 96 112 120 124 126 127 |
| 8 | 256 | 1 2 3 4 6 7 8 12 14 15 16 24 28 30 31 32 48 56 60 62 63 64 96 112 120 124 126 127 128 192 224 240 248 252 254 255 |
| 10 | 1024 | 1 2 3 4 6 7 8 12 14 15 16 24 28 30 31 32 48 56 60 62 63 64 96 112 120 124 126 127 128 192 224 240 248 252 254 255 256 384 448 480 496 504 508 510 511 512 768 896 960 992 1008 1016 1020 1022 1023 |

**Table 1.** Frequency multiplication factors $\mu_R = k/2^N$ realizable with the $\mu$-multiplier containing $N$ random flip-flops.

Frequency of an input signal is randomly multiplied by a constant number $0 < \mu_R < 1$. Using various combinations of $n$ and $m$ DRFFs one can obtain various multiplier values in the form $k/2^{(n+m)}$ where $1 \leq k \leq 2^{(n+m)}\text{-}1$. Unfortunately not all $k$'s in the range can be obtained. Table 1 summarizes all values of $k$ realizable by $N$=$n$+$m$ random flip-flops. Some values of the multiplication factor $\mu_R$ seem to be impossible to realize, for example 5/8, 9/16, 13/16, etc. In fact, only $N(N+1)/2$ values out of $2^N$-1 can actually be realized (e.g. 36 out of 255 for 8 flip-flops, 55 out of 1023 for 10 flip-flops, etc.), but the values are roughly uniformly distributed so this limitation may be acceptable in some applications where simplicity of the circuit is an important consideration. An exhaustive set of multiplication values can also be obtained but by a substantially more complex circuit explained in the next section.

## 9. Random frequency multiplier and synthesizer ($\lambda$-multiplier)

Frequency synthesis of periodic signals can be done by well known techniques, for example phase-locked loop (PLL) or direct digital synthesis (DDS). These enable generation of a set of frequencies separated by an arbitrary small increment in a desired frequency interval. These techniques are useful in a wide variety of applications from telecommunications to laboratory equipment. On the other hand, methods for systematic frequency synthesis of random pulse trains are virtually unknown.

Using random flip-flops, one is able to build a full-fledged random frequency synthesizer with frequency settable between arbitrary low and high frequencies in constant, arbitrary small increments. Such a device is analogous to usual PLL synthesized digital periodic pulse generators except that its output contains much more information. Namely, it delivers constant amount of entropy per pulse, whereas a periodic generator (ideally) delivers zero entropy (no information) per pulse. A realization of a random frequency synthesizer by use of the circuit named "$\lambda$-multiplier" is shown in Fig. 13. First, define two sub-circuits. A *random pulse splitter* (RPS) shown in Fig. 13a transmits a pulse from the input randomly to one of the outputs. No pulses are lost: any pulse entering the input IN ends up either exiting the output OUT1 or OUT2. A sole function of the delay circuit DLY shown in Fig. 13b is to delay pulses by exactly the same amount of time as RPS does. The schematic shown is correct if the DRFF is constructed as shown in Fig. 6. The lambda multiplier itself is shown in Fig. 13c. We assume a single RPT generator of frequency $f$ connected to the Input. The input pulses are randomly directed towards (mechanical or electrical) switches $S_i$ with probabilities $1/2^i$. Depending on the setting of switches, none, some or all pulses will be randomly chosen and conveyed to the Output, thus effectuating frequency multiplication by a factor $\lambda_R = k/2^N$ where $0 \leq k \leq 2^N$ and $k$ depends on the settings of the switches. This allows to generate a RPT with frequency in the range $[0, f]$ in constant steps of $f/2^N$. One should consider this particular realization of the $\lambda$–multiplier as being very resource-efficient because it requires only $N$ technologically "expensive" DRFFs to reach frequency step of $2^{-N}$ and $(N^2+3N)/2$ ordinary flip-flops which are technologically "cheap" and can be provided in abundance by use of conventional VLSI or reconfigurable technologies.

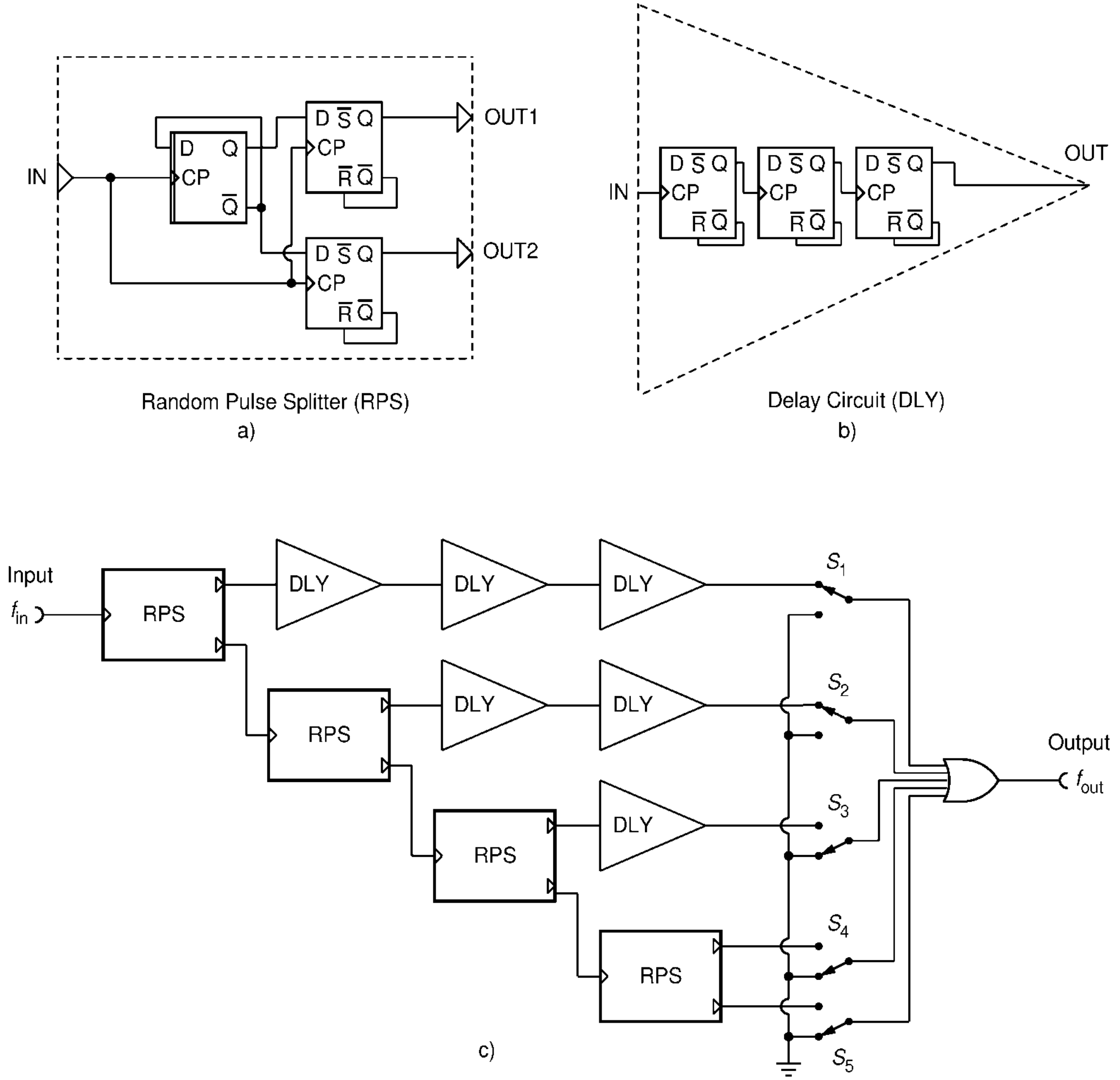

**Figure 13.** Lambda multiplier: a random frequency multiplier.

A random pulse train frequency generator that can synthesize RPT in an arbitrary range between $f_1$ and $f_2$ in steps of at most $\Delta f$ can now be built from a $\lambda$-multiplier, two RPT generators and a frequency summing circuit, as shown in Fig. 14. with the number of flip-flops (stages) $N$ chosen such that $f/2^N < \Delta f$. Frequency summing circuits are discussed in Sect. 11. Precise, quickly settable and stable random pulse train generators that would produce required fixed frequencies $f_{in} = f_2 - f_1$ and $f_1$ are described in [24] and [25].

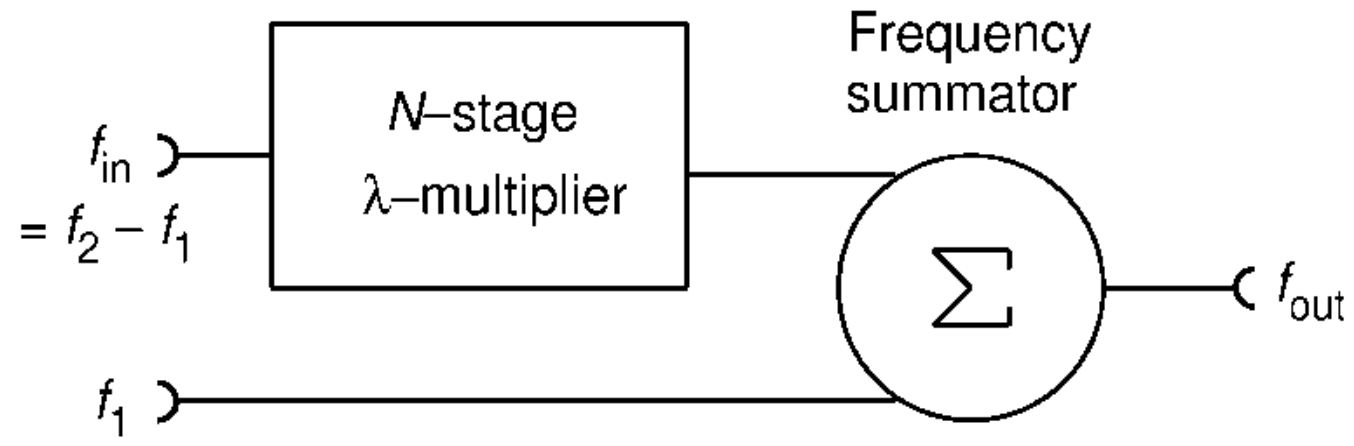

**Figure 14.** Synthesizer of random pulse trains of frequency in an arbitrary range [$f_1$, $f_2$] adjustable in arbitrary small steps of size $(f_2\text{-}f_1)/2^N$.

Random pulse generators with a precisely tunable frequency can be used in test equipment, research of random number generators, simulation and calibration of nuclear and radiation detectors, in secure authentication methods etc. [24], [26].

## 10. Random Programming Language Instructions

Should computer microprocessors be equipped with random logic circuits, such as random flip-flop, they would be able to execute random instructions, for example random branching instruction. Conventional branching instruction is used in a context similar to this:

```
IF {condition} THEN
 {block1}
ELSE
 {block2}
ENDIF
```

where *condition* is a Boolean-valued function, such as "A>B". If *condition* is fulfilled then program flow deterministically branches to the *block1* otherwise to *block2* after which it continues at the first instruction after ENDIF. Following the logical reversibility of computing [19] this (and in fact any) instruction can be inverted. To facilitate inversion, special universal programming languages have been formulated (e.g. R, PISA, LISP). This led to an automatic program inverter in LISP [27]. Invertible programming is important because it promises computing with reduced power dissipation - a problem that has stopped Moore's law advance in processor clock speed about year 2005. Probably the most devastating practical consequence of inverse programming is that any cryptographic primitive, such as AES, has its inverse. However, instructions relying on true randomness are non-invertible and therefore hold promise of safer cryptography [28]. A random branching instruction may have the following form:

```
RIF {prob} {condition} THEN
 {block1}
ELSE
 {block2}
ENDIF
```

where *prob* could be a real-valued probability of branching to *block1* if *condition* is TRUE, otherwise branching to *block2*. In a real-world implementation parameter *prob* could be compared to a single-precision float number which consists of 24 bit random mantissa and fixed 8 bit exponent and sign part. A 24 random flip-flop register could produce one such random number in a single processor clock – incomparably faster than any pseudo-random number algorithm executed in software, and on top of that truly random and non-invertible. A version of this would be random GOTO instruction that would branch to any of the labels in the list with equal probability:

RGOTO label1, label2, label3, ...

Another issue is the speed of random number production. In Monte Carlo programs with intensive use of random numbers their production in software can amount a significant portion of CPU resources. Having a specialized register consisting of random flip-flops would greatly improve CPU performance.

## 11. Computing with random pulse trains

An interesting notion of a computer based on random pulse trains [29], [30-31] makes use of various operations with random pulse trains. In this computation paradigm numerical values of variables and constants correspond to mean frequencies of RPTs. A number of circuits has been devised that can process this type of information [30-31]. Input and output values are frequencies of random signals. It possible to compute various mathematical functions at a very small hardware budget. We have already seen how multiplication of RPT's frequency with an arbitrary constant less than one can be performed by the $\lambda$-multiplier. Figure 15 shows three binary operations originally introduced in [30]. A single 2-input AND gate can perform exact multiplication of two frequencies as shown in Fig.15a.

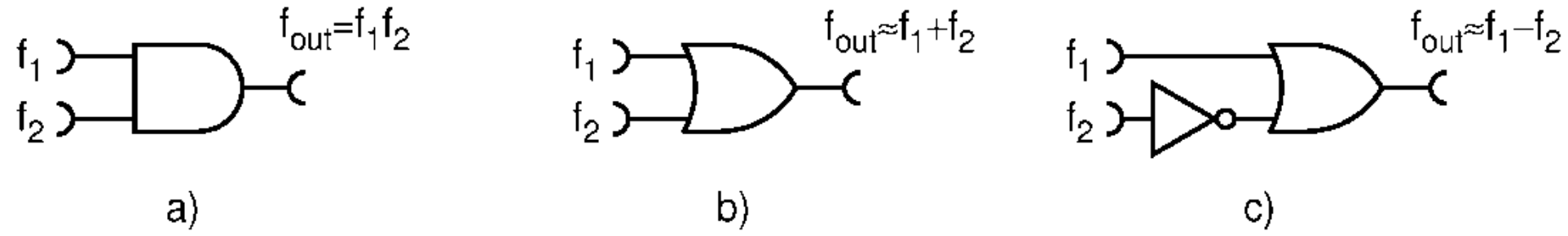

**Figure 15.** Realizations of elementary mathematical operations between frequencies of Poissonian random pulse trains. A single AND gate performs an exact multiplication (a), a single OR gate performs an approximate summation (b) while OR gate with one input preceded by NOT gate performs an approximate subtraction.

This works because the frequency of coincidences of two random pulse trains, one containing pulses of the width $w_1$ and the other of $w_2$ (usually they would be the same), and of respective frequencies $f_1$ and $f_2$ is given by:

$$f_{out} = (w_1 + w_2 + g)f_1 f_2 \tag{5}$$

where $g$ is a small correction due to the setup time of the AND gate. A 2-input OR gate will perform approximate summation (Fig. 15b) or subtraction (Fig. 15c) of frequencies of RPTs. The reason why summation (subtraction) is only approximate is because the input pulses have finite width and thus can overlap in which case two pulses are counted as one. The absolute value of the result is therefore smaller than the exact result. The accuracy can be improved by using shorter pulses but because of finite setting time of logic gates and because short pulses diminish signal from the multiplying circuit Eq. (5), pulses cannot be too short. Here, we propose a novel, edge-sensitive summing circuit shown in Fig. 16 which performs exact summation. The output appears divided by $2_R$ and can be restored to the usual RPT format by the frequency doubler described earlier (Fig. 10).

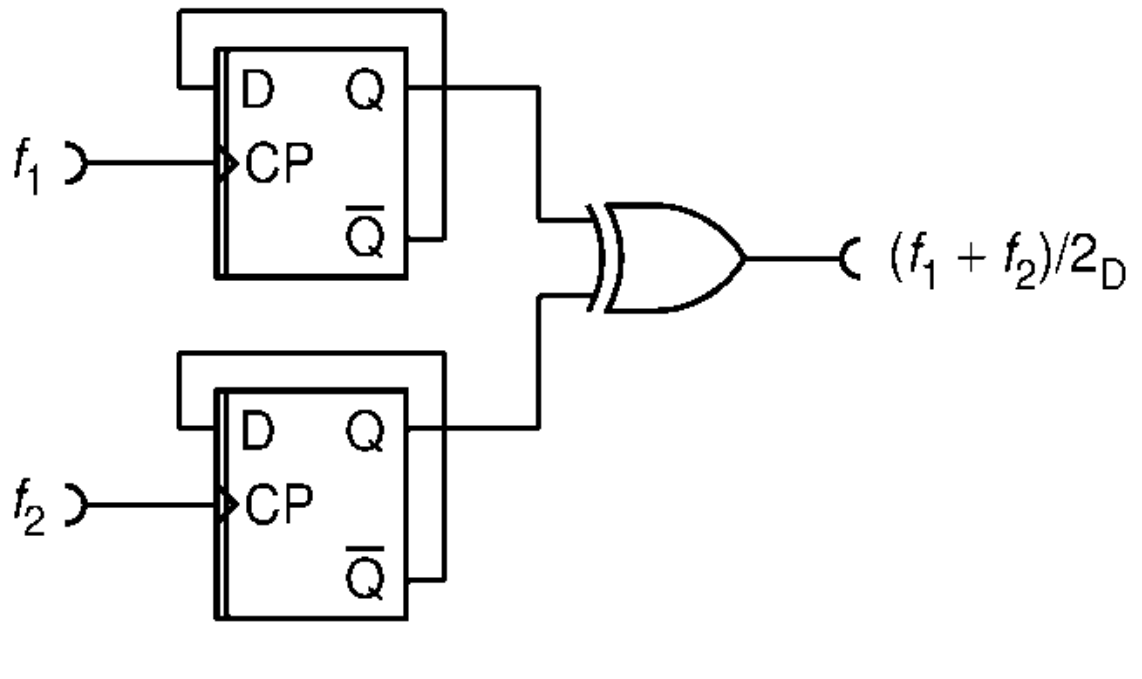

**Figure 16.** Frequency summing circuit which sums exactly frequencies two input signals. The output appears divided deterministically by 2, exactly as being divided by a single toggle flip-flop.

Simultaneous operation(s) on more than two variables can be done with almost no expense on hardware. For example, simultaneous multiplication of several numbers can be done by a single AND gate with appropriate number of inputs, each input connected to one RPT. Similarly, a summation of *n* RPTs can be done by the circuit similar to the one in Fig. 16 that would use *n* flip-flops connected to an *n*-input XOR. (An *n*-input XOR, of course, operates as a sum of all its inputs modulo 2.) Note that there is very little additional resources and no change in propagation delay by going from two to many numbers, which is to be contrasted to the digital calculation scheme where additional hardware and calculation time would be substantially larger. Having these basic operations (multiplication, summation, subtraction and multiplication by a constant) one can evaluate polynomials and thus many elementary mathematical functions (via Taylor's expansion). However, as opposed to frequency of a periodic signal, frequency of an RPT cannot be simply multiplied by a number greater than one without actually adding the required amount of entropy. An entropy conservation principle can be formulated in that respect [32]. That is why the RPT computing paradigm is best suited for numerical values between 0 and 1, and therefore doing math in a RPT computer may involve careful normalization. It is also known how to divide two frequencies (the fourth elementary operation, along with multiplication, summing and subtracting) but so far it is not known how to do it in a direct way, but rather only via a (slow) frequency-to-amplitude conversion [30].

An RTP computer possesses and inherent immunity to accidental errors and noise due to the stochastic nature of the computational substance (pulse trains). A small addition or subtraction of pulses due to hardware errors or noise would give only small effect on the result while a single bit flip in digital computing on average leads to a dramatically different result of calculation. This is probably why RPT computer was first designed for use in airplane steering mechanism [38]. Its precision is no match for modern digital computers but with the advantage of resilience to harsh environment and small requirement for resources it may be interesting in some applications. This computing paradigm is peculiar because it seems that randomness itself does the calculation: replace RPTs with periodic signals and calculation is gone.

## 12. Gaussian white noise generator

Gaussian white noise is a very useful resource for various techniques such as: real time speech and music synthesis, measuring and testing of audio equipment and radio communication circuits. Random logic can be used quite elegantly for production of analog white noise with exceptionally flat power density spectrum. Circuit shown in Fig. 17, driven by the periodic oscillator of frequency $f_{osc}$, generates at its output Gaussian white noise whose frequency spectrum envelope is nearly flat from 0 Hz till some portion of $f_{osc}$. The envelope of the power density is given by:

$$\frac{dP}{df} = \left( \frac{\sin\left(\pi \frac{f}{f_{clock}}\right)}{\pi \frac{f}{f_{clock}}} \right)^2 \tag{6}$$

The power density continuously drops down falling only -0.1dB at 0.12fclock, -1dB 0.26fclock at and half of the power (-3dB) at 0.44fclock. If faster drop at the end of the spectrum is desired, one could pass the output through an active analog low-pass filter. In that case, the flatness of the spectrum is also affected by the transfer function of the filter. As is well known in the art, for maximum flatness in the pass region one should use the *n*-th order Butterworth filter, whereas for the maximum steepness in the stop region and minimum ripples in the pass region one should use the Chebyshev type of low-pass filter [33]. Digital filtering offers further possibilities to shape the power density virtually at will.

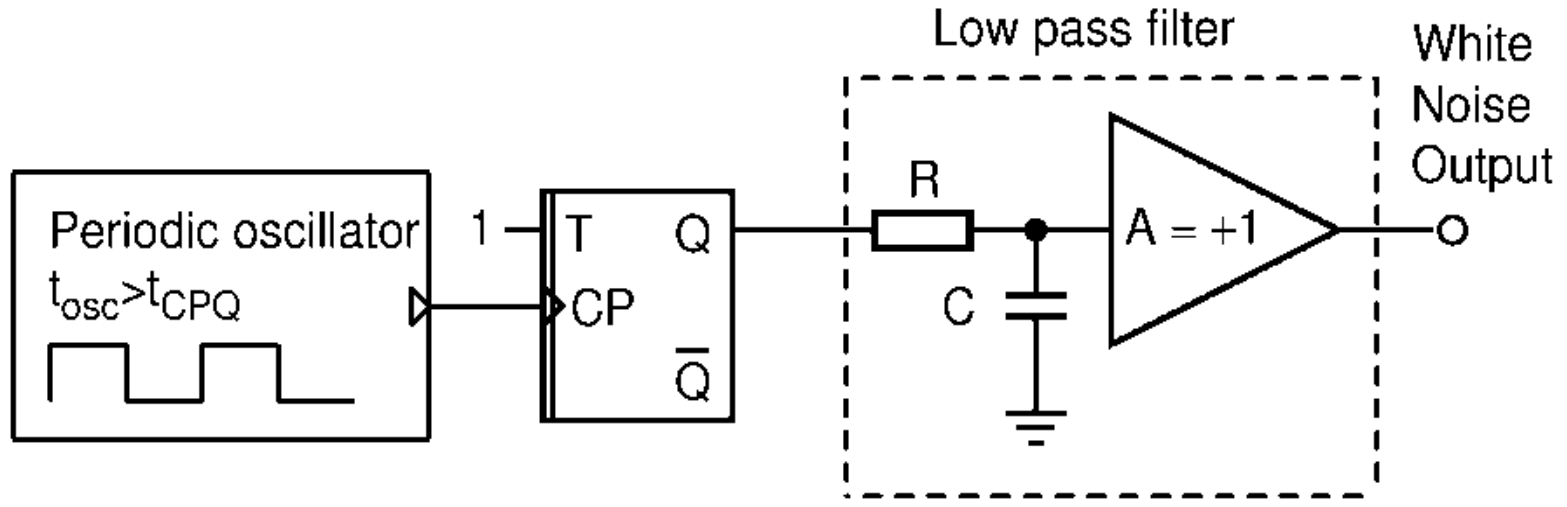

**Figure 17.** A practical realization of analog Gaussian white noise generator by means of a single random flip-flop.

A detailed mathematical treatment of this type of white noise generator and its spectrum is given in [34] and an improved version in [35] while suitable low-pass filters are explained for example in [33]. If the pseudorandom source was used, the spectrum would consist of equidistantly spaced spikes (excluding 0Hz) rather than being truly flat. The comb-like spikes in the pseudo-random noise come from the finite period of the pseudo-random number generator, whereas the period of a non-deterministic generator is infinite and consequently the spectrum is continuous. An interesting characteristic of the analog noise generator presented here is that the spectrum is extremely flat at the low frequency end and extends all the way to 0 Hz, which is to be contrasted to various semiconductor sources, such as Zener diodes or bipolar transistors whose spectrums suffer from 1/f rise in that region.

## 13. Binary channel with noise

In information-theoretic secure cryptographic protocols based on noisy channels [36-37] the crucial resource is a physical device called "symmetric binary channel with noise", which flips the data bit with a predetermined probability $p_{err}$, known also as *bit error probability*. Such a device effectively adds a precise, predetermined quantity of non-deterministic noise (entropy) to the stream of digital data bits. Circuit shown in Fig. 18 is a practical realization of a binary noisy channel using previously described $\lambda$-multiplier. We assume that the data bits at the Data Input appear in synchronization with the positive going edges of the Strobe. The data bits and the random sequence from the divider's output are synchronized by means of the pair of D-type flip flops and then XOR-ed to obtain a bit flip with probability $p_{err} = \lambda$.

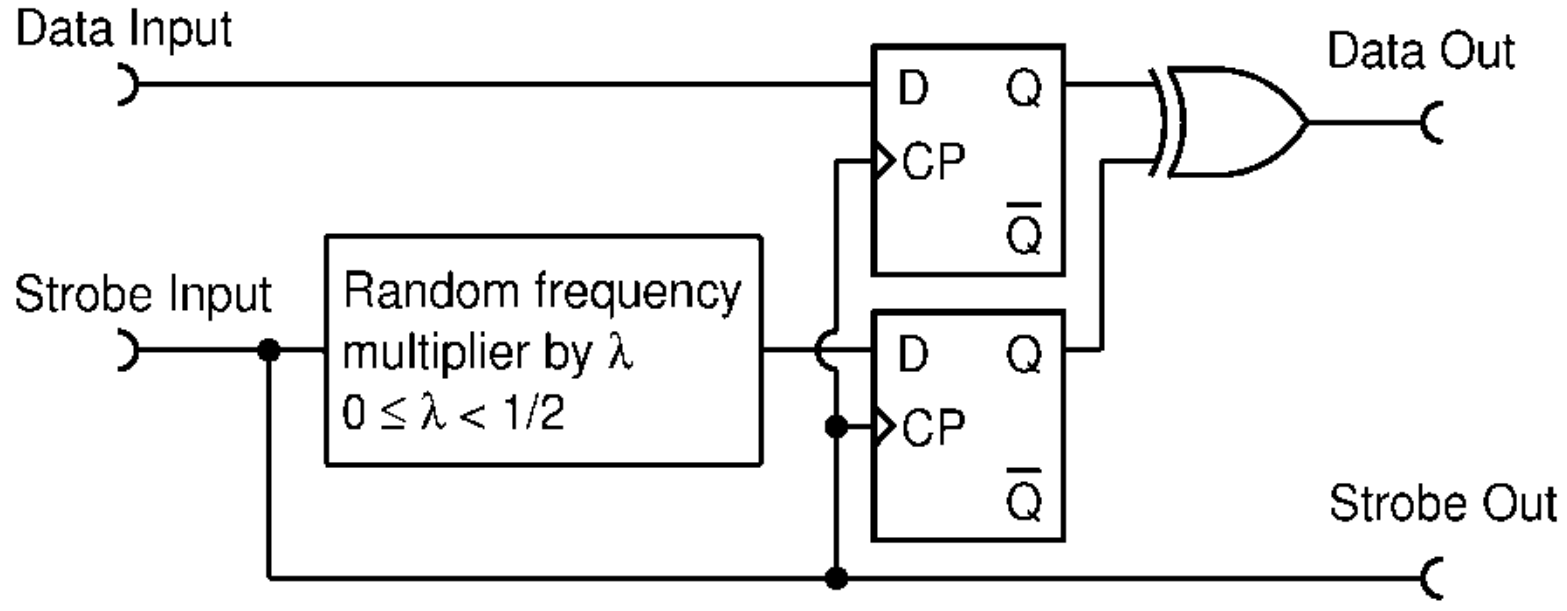

**Figure 18.** Realization of symmetric binary channel with noise by means of the random divider circuit. Input data bits are either copied to the output with probability (1 - $\lambda$) or inverted with the probability $\lambda$.

For binary channels with noise considered in information theory bit error probability is usually a small number absolutely limited by $0 \le p_{err} < 0.5$ which can be accomplished by the $\mu$-multiplier ore the $\lambda$-multiplier.

## 14. Dice thrower

And finally some fun. Circuit in Fig. 19 after each request generates one random number between 1 and 6 (in binary or BCD notation) thus emulating throws of a perfectly fair dice. What happens is that upon a request signal on the input "Request" the device produces one random number in the range 0-7. If it happens to produce 0 or 7 then the periodic oscillator sends additional requests until the random number happens to fall between 1 and 6. New requests are not possible until the last one is completed which is signaled by "Ready" going HIGH. Note that this particular circuit is permutation symmetric in the three flip-flops meaning that it is irrelevant which binary value is assigned to which output. Moreover, the assignment can be changed at will and at any time - the circuit will still produce sequence of absolutely fair and uncorrelated random throws! This "dice" is unpredictable: collection of all throws made so far contains zero information about the outcome of the next throw. This example illustrates usability of RFFs in hazard games, online gambling and gaming automata.

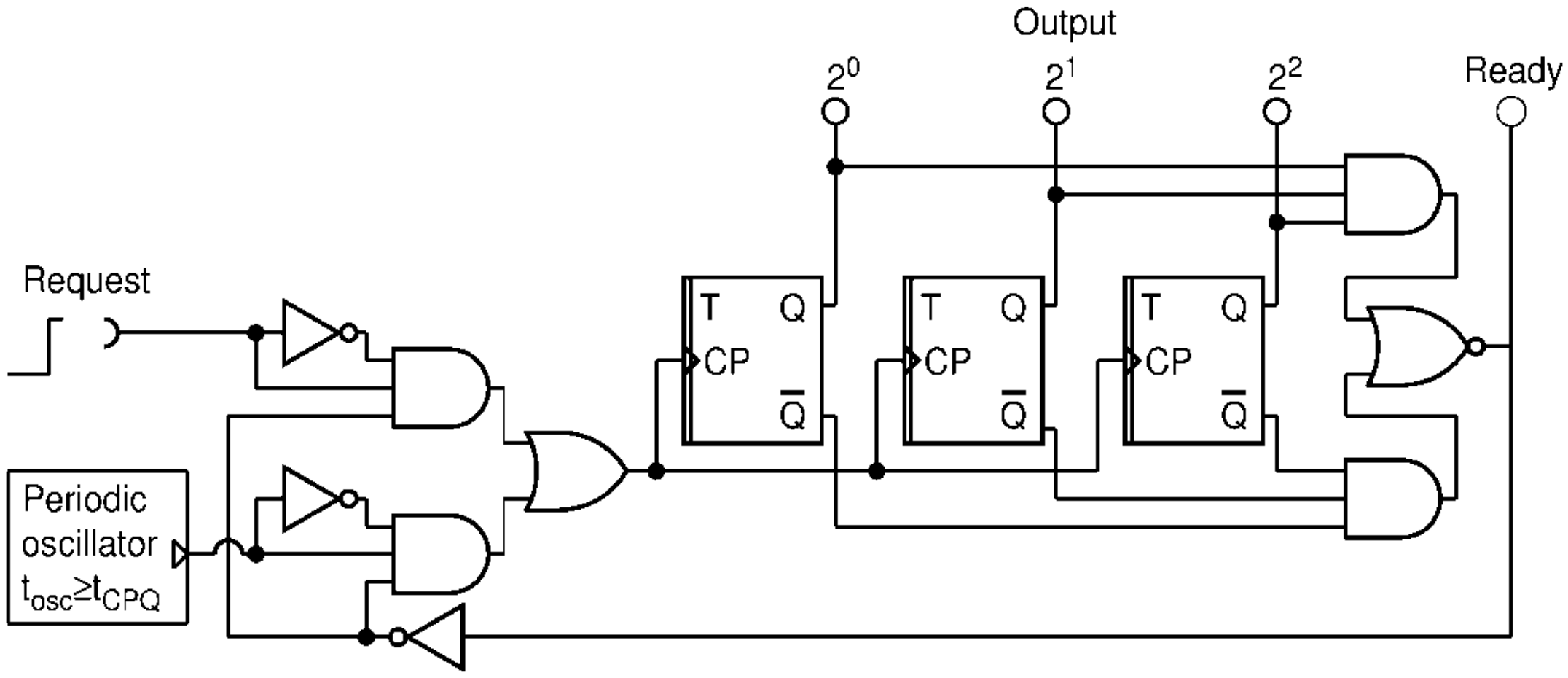

**Figure 19.** A random logic circuit which generates equiprobably one random binary number in the range of 1 to 6 upon each request, thus mimicking throws of a perfectly fair dice.

## Conclusion

We introduce a new non-sequential logic element, the random flip-flop, and discuss its applications in random numbers, random counters, randomness preserving frequency dividers, random pulse train frequency synthesis, probabilistic Turing computers, random pulse train computers, analog noise generators, binary channels with noise and gambling automata. We also show how to build a D-type random flip-flop based on single photon detection of random light, using existing state of the art.